\documentclass{jkas}

\def\year{2019} 
\def\month{March} 
\def\volume{00} 
\def\issue{0} 
\def\beginpage{1} 
\def\endpage{99} 
\def\received{March 15, 2019} 
\def\accepted{March 15, 2019} 
\date{Received \received; accepted \accepted}

\usepackage{graphicx}
\usepackage{amsmath}

\usepackage[varg]{txfonts}

\usepackage{bm}
\usepackage{mathtools}
\usepackage{natbib}
\usepackage{lineno}
\usepackage{hyperref}

\title{
Forecast of Daily Major Flare Probability Using Relationships between Vector Magnetic Properties and Flaring Rates
}

\author[1]{Daye Lim}
\author[1]{Yong-Jae Moon}
\author[2]{Jongyeob Park}
\author[1]{Eunsu Park}
\author[1,3]{Kangjin Lee}
\author[4]{Jin-Yi Lee}
\author[2]{Soojeong Jang}

\affil[1]{School of Space Research, Kyung Hee University,
1732, Deogyeong-daero, Giheung-gu, Yongin-si,
Gyeonggi-do 17104, Korea; \email{moonyj@khu.ac.kr}}
\affil[2]{Korea Astronomy and Space Science Institute,
776, Daedeokdae-ro, Yuseong-gu, Daejeon,
34055, Korea}
\affil[3]{Electronics and Telecommunications Research Institute,
218, Gajeong-ro, Yuseong-gu, Daejeon,
34129, Korea}
\affil[4]{Department of Astronomy \& Space Science, Kyung Hee University,
1732, Deogyeong-daero, Giheung-gu, Yongin-si,
Gyeonggi-do 17104, Korea}

\def\corrauthor{
Y.-J. Moon (\email{moonyj@khu.ac.kr})
}

\def\runningauthor{
Lim et al.
}

\def\runningtitle{
Flare forecasting model
}

\def\keywords{
Sun: activity -- Sun: flares -- Sun: magnetic fields
}

\def\abstracttext{
We develop forecast models of daily probability of major flares (M- and X-class) based on empirical relationships between photospheric magnetic parameters and daily flaring rates from May 2010 to April 2018. In this study, we consider ten magnetic parameters characterizing size, distribution, and non-potentiality of vector magnetic fields from \textit{Solar Dynamics Observatory} (\textit{SDO})/Helioseismic and Magnetic Imager (HMI) and \textit{Geostationary Operational Environmental Satellites} (\textit{GOES}) X-ray flare data. The magnetic parameters are classified into three types: the total unsigned parameters, the total signed parameters, and the mean parameters. We divide the data into two sets chronologically: 70\% for training and 30\% for testing.  The empirical relationships between the parameters and flaring rates are used to predict flare occurrence probabilities for a given magnetic parameter value. Major results of this study are as follows. First, major flare occurrence rates are well correlated with ten parameters having correlation coefficients above 0.85. Second, logarithmic values of flaring rates are well approximated by linear equations. Third, the total unsigned and signed parameters achieved better performance for predicting flares than the mean parameters in terms of verification measures of probabilistic and converted binary forecasts. We conclude that the total quantity of non-potentiality of magnetic fields is crucial for flare forecasting among the magnetic parameters considered in this study. When this model is applied for operational use, this model can be used using the data of 21:00 TAI with a slight underestimation of 2 - 6.3\%.
}

\begin{document}
\jkashead 


\section{Introduction\label{sec:intro}}

Solar flares rapidly release a tremendous amount of energy as the form of electromagnetic radiation, high energy particles and shock waves, which result in space weather hazards. In space age, precautions against economic risk by predicting solar flare occurrence are very essential \citep{Tsurutani05, Schwenn06, Bocchialini18}.

Most solar energetic events such as solar flares explode out of active regions (ARs), which are areas of complex and intense photospheric magnetic fields. Thus, characteristics of solar ARs are considered to closely be related to solar flares. Several different types of AR information have been used for solar flare forecasts. Many studies have considered morphological characteristics of ARs such as size, shape and complexity using Mount Wilson magnetic classification and McIntosh classification \citep{Hale19, McIntosh90, Bornmann94, Sammis00, Gallagher02, McAteer05, Qahwaji07, Li08, Colak09, Bloomfield12, Lee12, Li13, Lee16, McCloskey16, Shin16}. In addition, various magnetic parameters characterizing distribution and non-potentiality of ARs from magnetograms have been studied for flare forecasting \citep{Leka03a, Leka03b, Cui06, Leka07, Barnes07, Schrijver07, Yu09, Yuan10, Falconer11, Ahmed13, Huang13, Falconer14, Bobra15, Barnes16, Nishizuka17, Raboonik17, Liu17, Nishizuka18, Huang18, Leka18, Park18}. The magnetic parameters have been also calculated by magnetohydrodynamic (MHD) simulations and used for predicting flares \citep{Aulanier13, Guennou17, Toriumi17}.

There have been many studies on flare probability depending on AR's characteristics and flare probability forecasting models. \citet{Giovanelli39} examined probabilities of an solar eruption in relation to sunspot group's size, type, and development. \citet{Gallagher02} developed a flare prediction system using McIntosh classification, which gives daily flare probabilities based on Poisson statistics. \citet{Leka03a} considered magnetic parameters such as the vertical current, the current helicity, the twist parameter $\alpha$, and the magnetic shear angles for discriminating flaring and flare-quiet ARs, which had been examined in a series of papers \citep{Leka03b, Barnes06, Leka07}. In order to parameterize these magnetic parameters, they first used vector magnetograms from Imaging Vector Magnetograph (IVM) at the University of Hawai'i Mees Solar Observatory. As a forecast result of their studies, \citet{Barnes07} provided flare probabilities using Bayes's theorem. \citet{Falconer11} found an empirical relationship between flare event rates and a proxy of magnetic free energy based on line-of-sight magnetograms from \textit{Solar and Heliospheric Observatory} (\textit{SOHO}; \citealt{Domingo95})/Michelson Doppler Imager (MDI; \citealt{Scherrer95}). They predicted flare probabilities from a power law relationship between M- and X-class flare occurrence rates and the free magnetic energy proxy in order to forecast flares. \citet{Bloomfield12} determined the Poisson flare probabilities from McIntosh sunspot classes. Among a variety of forecast verification scores, they proposed true skill statistic (TSS) as a standard score for comparing between flare forecasts. They also presented optimum thresholds to convert probabilities into binary predictions and forecast verification measures using these thresholds. \citet{Lee12} also reported flare occurrence Poisson probabilities of McIntosh classification. They found that these flare probabilities tend to increase with AR's area and flaring probabilities for increasing AR's area are higher than those for steadying and decreasing area.

Recently, \textit{Solar Dynamics Observatory} (\textit{SDO}; \citealt{Pesnell12}) was launched in 2010 and Helioseismic and Magnetic Imager (HMI), which is one of three instruments on the \textit{SDO}, provides full-disk photospheric vector magnetic fields with 12 minutes cadence \citep{Scherrer12, Schou12, Hoeksema14}.

The HMI team developed a set of derivative data called Space-weather HMI Active Region Patches (SHARPs) data \citep{Bobra14}. These data contain automatically identified HMI Active Region Patches (HARPs) and magnetic parameters which summarize the size, distribution, and non-potentiality of vector magnetic fields in each HARP, and these parameters have been adapted from numerous studies \citep{Leka03a, Schrijver07, Fisher12}. SHARP data have been applied to machine learning algorithms for binary flare forecasting \citep{Bobra15, Nishizuka17, Raboonik17, Liu17, Nishizuka18}. The probabilistic forecasting have been performed by \citet{Kontogiannis17} and \citet{Leka18} using Bayesian probabilities. Although there have been many categorical forecasts using vector magnetic fields, probabilistic forecasts have been rarely considered. SHARP magnetic parameters are also ranked by performance of discriminating between flaring events and non-flaring events based on machine learning algorithm \citep{Bobra15, Liu17}.

In this paper, we study empirical relationships between photospheric SHARP magnetic parameters and daily major flaring rates. Furthermore, we will develop forecast models of daily probability of major flares (M- and X-class) based on these empirical relationships. We will also identify the rank of SHARP magnetic parameters examining the performance of the models developed from the parameters.
In general, the occurrence probability could present more continuous information on potentiality of flare occurrence than the binary (flaring/non-flaring) forecasts. Furthermore, the probability forecasts can be converted into binary forecasts using proper threshold values \citep{Colak09, Crown12, Bloomfield12, Park17, Murray17, Leka18}.

The paper is organized as follows. Section \ref{sec:data} deal with detailed description of the data. The empirical relationships using our model is explained in detail in Section \ref{sec:model}.
Forecasting models and their results are given in Section \ref{sec:results}. A summary and discussion are presented in Section \ref{sec:summary}.

\section{DATA AND ANALYSIS} \label{sec:data}
\subsection{\textit{Geostationary Operational Environmental Satellites (GOES) X-ray flares}}

\textit{GOES} have measured solar X-rays in the passbands of 1 - 8 \AA\ and 0.5 - 4 \AA. Solar X-ray flares are classified according to the peak flux of X-rays with wavelength bands 1 to 8 \AA\ as measured by \textit{GOES}.
We use \textit{GOES} major (M- and X-class) X-ray flare data from May 2010 to April 2018 and their locations are identified by the Lockheed Martin Solar and Astrophysics Laboratory (LMSAL)\footnote{\url{https://www.lmsal.com/solarsoft/latest_events_archive.html}}. Our data include 448 M-class flares and 27 X-class flares.

\subsection{\textit{SDO}/HMI and magnetic parameters}

SHARP magnetic parameters have been used for flare forecasting \citep{Bobra15, Liu17} based on machine learning algorithm. Among the SHARP parameters, we just consider ten parameters which have linear Pearson correlation coefficients (CCs) between these and flaring rates higher than 0.85 described in Section \ref{sec:model}. The ten parameters are classified into three types: the total unsigned parameters (TOTUSJH $H_{C_{\text{total}}}$, TOTUSJZ $J_{{z\text{,total}}}$, TOTPOT $\rho_{\text{tot}}$, and USFLUX $\Phi$), the total signed parameters (SAVNCPP $J_{{z\text{,sum}}}$ and ABSNJZH $H_{{C\text{,abs}}}$), and the mean parameters (MEANPOT $\overline{\rho}$, SHRGT45 $A_{\text{shear}}$, MEANSHR $\overline{\Gamma}$, and MEANGAM $\overline{\gamma}$). The description and formula of these parameters are listed in Table \ref{tab:tbl1}.

\begin{table*}[]
\caption{The description and formula of ten SHARP magnetic parameters}
\label{tab:tbl1}
\centering
\resizebox{\textwidth}{!}{%
\begin{tabular}{lll}
\hline
\hline
Keyword & Description & Formula \\ \hline
TOTUSJH & Total unsigned current helicity & $H_{C_{\text{total}}}$ $=$ $\sum$ $|$$B_{z}$$\cdot$$J_{z}$$|$ \\
TOTUSJZ & Total unsigned vertical current & $J_{{z\text{,total}}}$ $=$ $\sum$ $|J_{z}|$ $dA$ \\
TOTPOT & Total photospheric magnetic free energy density & $\rho_{\text{tot}}$ $=$ $\sum$ ($\bm{B}_{obs}$-$\bm{B}_{pot})^2$ $dA$ \\
USFLUX & Total unsigned magnetic flux & $\Phi$ = $\sum$ $|B_{z}|$ $dA$ \\
SAVNCPP & Sum of the net current emanating from each polarity & $J_{{z\text{,sum}}}$ $=$ $|$$\sum^{B_{z}^{+}}$ $J_{z} dA$$|$+$|$$\sum^{B_{z}^{-}}$ $J_{z} dA$$|$ \\
ABSNJZH & Absolute value of the net current helicity & $H_{{C\text{,abs}}}$ $=$ $|$$\sum$$B_{z}$$\cdot$$J_{z}$$|$ \\
MEANPOT & Mean photospheric magnetic free energy density & $\overline{\rho}$ $=$ $\frac{1}{N}$ $\sum$ $(\bm{B}_{obs}-\bm{B}_{pot})^2$ \\
SHRGT45 & Fractional area with shear $> 45^\circ$ & $A_{\text{shear}}$ $=$ Area with shear $> 45^\circ /$ HARP area \\
MEANSHR & Mean shear angle & $\overline{\Gamma}$ $=$ $\frac{1}{N}$ $\sum$ arccos $(\frac{\bm{B}_{obs} \cdot \bm{B}_{pot}}{|B_{obs}||B_{pot}|})$ \\
MEANGAM & Mean angle of field from radial & $\overline{\gamma}$ $=$ $\frac{1}{N}$ $\sum$ arctan $(\frac{B_{h}}{B_{z}})$ \\ \hline
\end{tabular}%
}
\tabnote{Constant terms are omitted.}
\end{table*}

We use 00:00 TAI definitive HARPs in cylindrical equal area (CEA) coordinates (hmi.sharp\_720s\_cea data series) when their longitudes are within $\pm$ 60 degrees of the central meridian and corresponding ten magnetic parameters from the Joint Science Operations Center (JSOC)\footnote{\url{http://jsoc.stanford.edu/}}. According to \citet{Hoeksema14}, the number of high-confidence pixels in SHARP data decreases significantly beyond $\pm$ 60 degrees of the central meridian. In this study, a HARP is regarded as an unit of area to occur flares.


\section{Empirical Relationships between Magnetic Parameters and Solar Major Flare Occurrence Rates} \label{sec:model}

To develop a solar flare occurrence probability forecasting model, we need relationships between ten magnetic parameters and major flare occurrence rates. For identifying these relationships, we need a data set of major flare occurrence history and corresponding ten parameter values for each HARP.

\subsection{Flare identification}

We identify each HARP at 00:00 TAI that produced one or more major flares within a day by using the flare locations which are corrected for differential rotation rates. When a flare event is located in a HARP's box, the flare event is considered to occur from that HARP. We assume that all flare events are independent of one another.

\subsection{Data set}

Our data sets are divided into two sets (training and test) in chronological order. 70\% of the data, HARPs from 1 May 2010 to 20 April 2015 including the ascending and maximum phase of the solar cycle (SC) 24, are used for finding a relationship between parameters and flaring rates.  And 30\% of the data, HARPs from 21 April 2015 to 30 April 2018 including the part of the descending phase of SC 24, are used for testing it. Accordingly, the training data consist of 11040 samples (different 1889 HARPs) and the test data consist of 4724 samples (different 898 HARPs). The training data sample consist of 224 event samples and 10816 non-event samples. The test data sample consist of 38 event samples and 4686 non-event samples. The sample ratio is unbalanced as well as previous major flare forecasting models, because major flares are rare events and the solar cycle 24 is unprecedented quiet.

\subsection{Major flare occurrence rates as a function of magnetic parameters}

\begin{figure} \centering
\resizebox{\hsize}{!}{\includegraphics{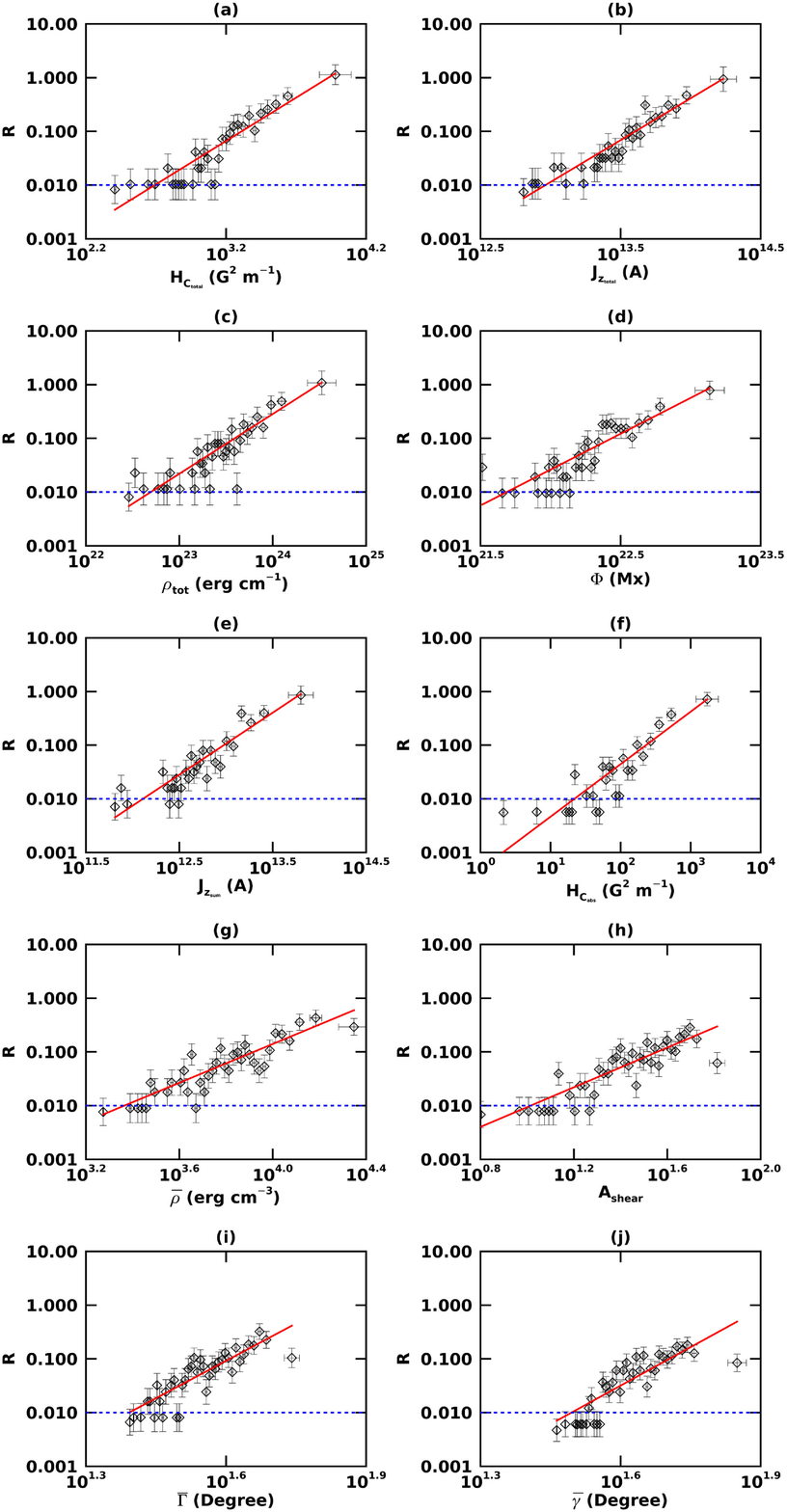}}
\caption{Major flare occurrence rates versus ten magnetic parameters in log-log scale. The occurrence rate of each group is shown with diamond. The vertical error bars represent the uncertainty of the occurrence rate of each group ($\sigma = \sqrt{\frac{R(1-R)}{N}}$). The horizontal error bars represent the root mean square error (RMSE) between parameter values of each group and its mean value. The red line is corresponding to the power law fitting. The data which have flaring rate above 0.01 (blue dashed line) are used in the fitting. \label{fig:f1}}
\end{figure}

We want to identify relationships between ten magnetic parameters and major flare occurrence rates. For each parameter, we divide our data into 50 subgroups in which there is the equal number of HARPs. Then, we determine the average parameter value and the number of major flares within a day from each daily HARP in each subgroup. Each mean major flare occurrence rate ($R_{i}$) of $i$-th group ($G_{i}$) is given by
\begin{equation}\label{eq:eq1}
R_{i} = \frac{\#\ \text{of major flares of}\ G_{i}}{\#\ \text{of HARPs of}\ G_{i}}.
\end{equation}
In order to find relationships, these rates are plotted as a function of each parameter in log-log scales. As \citet{Falconer11} presented a power law function of major flare occurrence rates and their parameter from line-of-sight magnetic fields, our models also considers the power law function of each parameter as shown in Figure \ref{fig:f1}. The occurrence rates range from 0.001 to about 1 for all parameters.
The flaring rates are only considered above 0.01 for obtaining more accurate fitting functions as \citet{Falconer11} did. The fitting function is given by the following
\begin{equation}\label{eq:eq2}
\log (R) =  a\log (x) + b,
\end{equation}
where $R$ is a mean flare occurrence rate, $x$ is a mean parameter value of a group, $a$ is a power law slope, and $b$ is a fitting constant. To examine the dependence of the binning size, we also consider three cases of binning size $=10, 20, 100$. Power law functions are well fitted with data for all four cases and their differences are very small. Thus, we use the power law function with binning size $=50$ and its fitting coefficients and uncertainties of ten SHARP parameters are shown in Table \ref{tab:tbl2}.

\begin{table*}[]
\caption{Correlation coefficients (CCs) between major flare occurrence rates and ten magnetic parameters, their fitting coefficients, and root mean square errors (RMSEs) between flaring rates and fitting lines}
\label{tab:tbl2}
\centering
\begin{tabular}{ccccccc}
\hline
\hline
Parameter & CC & $a$ & $b$ & RMSE \\ \hline
$H_{C\text{,total}}$ & $0.91$ & $1.61 \pm 0.003$ & $-6.34 \pm 0.012$ & $0.25$ \\
$J_{z\text{,total}}$ & $0.95$ & $1.56 \pm 0.003$ & $-22.21 \pm 0.045$ & $0.17$ \\
$\rho_{\text{tot}}$ & $0.87$ & $1.11 \pm 0.002$ & $-27.25 \pm 0.058$ & $0.27$ \\
$\Phi$ & $0.86$ & $1.34 \pm 0.004$ & $-30.95 \pm 0.08$ & $0.27$ \\
$J_{{z\text{,sum}}}$ & $0.90$ & $1.16 \pm 0.002$ & $-16.0 \pm 0.029$ & $0.24$ \\
$H_{{C\text{,abs}}}$ & $0.86$ & $0.98 \pm 0.002$ & $-3.31 \pm 0.004$ & $0.32$ \\
$\overline{\rho}$ & $0.90$ & $1.81 \pm 0.006$ & $-8.1 \pm 0.023$ & $0.22$ \\
$A_{\text{shear}}$ & $0.88$ & $1.85 \pm 0.006$ & $-3.87 \pm 0.009$ & $0.24$ \\
$\overline{\Gamma}$ & $0.86$ & $4.64 \pm 0.017$ & $-8.46 \pm 0.026$ & $0.24$ \\
$\overline{\gamma}$ & $0.86$ & $4.78 \pm 0.013$ & $-9.14 \pm 0.021$ & $0.27$ \\ \hline
\end{tabular}
\end{table*}

In Table \ref{tab:tbl2}, the CCs between flaring rates and ten SHARP parameters are high (above 0.85), implying that they are well correlated with each other. Among the ten parameters, the total unsigned vertical current has the minimum root mean square error (RMSE) between the power law fitting function and its original value (Figure \ref{fig:f1}a and \ref{fig:f1}b).

\section{Forecast Models and Their Verification} \label{sec:results}

\subsection{Forecast models based on empirical relationships between parameters and flaring rates}

We develop forecasting models of daily probability of major flares using the relationships between ten magnetic parameters and daily flaring rates. For a given $x$, we can predict a major flare occurrence rate from the empirical fitting functions. From these predicted rates, we calculate flare probabilities using the Poisson distribution \citep{Wheatland00, Moon01, Gallagher02, Bloomfield12, Lee12}. The probability ($P$) of occurring at least one flare in a day is given by
\begin{equation}\label{eq:eq3}
P=1-\text{exp}(-R).
\end{equation}

\subsection{Verification measures }

\subsubsection{Verification measures of probability}

Forecast models represent verification measures which are a single number measuring forecast performance. We consider the mean squared error (MSE), the Brier skill score (BSS), and the reliability plots. These measures have been used for verifying the performance of probabilistic forecast models \citep{Wheatland05, Barnes07, Barnes16}. The MSE, which is a measure of accuracy, is given by
\begin{equation}\label{eq:eq4}
\text{MSE}=\frac{1}{N}\sum_{i=1}^{N} (P_{i}-O_{i})^2,
\end{equation}
where $P_{i}$ is the predicted probability and $O_{i}$ is the observation that events occurred ($O_{i} = 1$) or did not occur ($O_{i} = 0$). When the perfect forecast occur, the MSE is 0. The BSS, which represents the relative skill compared to the model using the climatological event rate during the testing interval, is given by
 \begin{equation}\label{eq:eq5}
\text{BSS}=\frac{\frac{1}{N}\sum_{i=1}^{N} (P_{i}-O_{i})^2-\frac{1}{N}\sum_{i=1}^{N} (\overline{O_{i}}-O_{i})^2}{0-\frac{1}{N}\sum_{i=1}^{N} (\overline{O_{i}}-O_{i})^2}.
\end{equation}
When the perfect forecast occur, BSS is 1, and "no-skill" as compared to the climatological forecast results in 0. The reliability plots are observed occurrence rates against predicted occurrence rates. When the perfect forecast occur, the all points in the reliability plot lie on the diagonal line.

From the test samples, we calculate the MSE and BSS of ten forecast models, which are listed in Table \ref{tab:tbl3}. The total unsigned current helicity has the best performance in terms of both MSE and BSS. In view of the BSS, the total unsigned parameters have higher value than the signed and mean parameters; more specifically, $H_{C\text{,total}} > H_{{C\text{,abs}}}$, $J_{z\text{,total}} > J_{{z\text{,sum}}}$, and $\rho_{\text{tot}} > \overline{\rho}$.

\begin{table}[]
\caption{Verification measures of probabilistic forecasts: mean sqaured error (MSE) and Brier skill core (BSS) described in the text}
\label{tab:tbl3}
\centering
\begin{tabular}{ccc}
\hline
\hline
Parameter & MSE (Perfect = 0) & BSS (Perfect = 1) \\ \hline
$H_{C\text{,total}}$ & $0.006 \pm 0.0001$ & $0.22 \pm 0.005$ \\
$J_{z\text{,total}}$ & $0.007 \pm 0.0001$ & $0.17 \pm 0.004$ \\
$\rho_{\text{tot}}$ & $0.007 \pm 0.0001$ & $0.12 \pm 0.007$ \\
$\Phi$ & $0.007 \pm 0.0001$ & $0.12 \pm 0.004$ \\
$J_{{z\text{,sum}}}$ & $0.007 \pm 0.0001$ & $0.14 \pm 0.007$ \\
$H_{{C\text{,abs}}}$ & $0.007 \pm 0.0001$ & $0.15 \pm 0.007$ \\
$\overline{\rho}$ & $0.01 \pm 0.0001$ & $-0.26 \pm 0.014$ \\
$A_{\text{shear}}$ & $0.009 \pm 0.0001$ & $-0.15 \pm 0.009$ \\
$\overline{\Gamma}$ & $0.009 \pm 0.0001$ & $-0.18 \pm 0.01$ \\
$\overline{\gamma}$ & $0.009 \pm 0.0001$ & $-0.14 \pm 0.007$ \\ \hline
\end{tabular}
\end{table}

\begin{figure} \centering
\resizebox{\hsize}{!}{\includegraphics{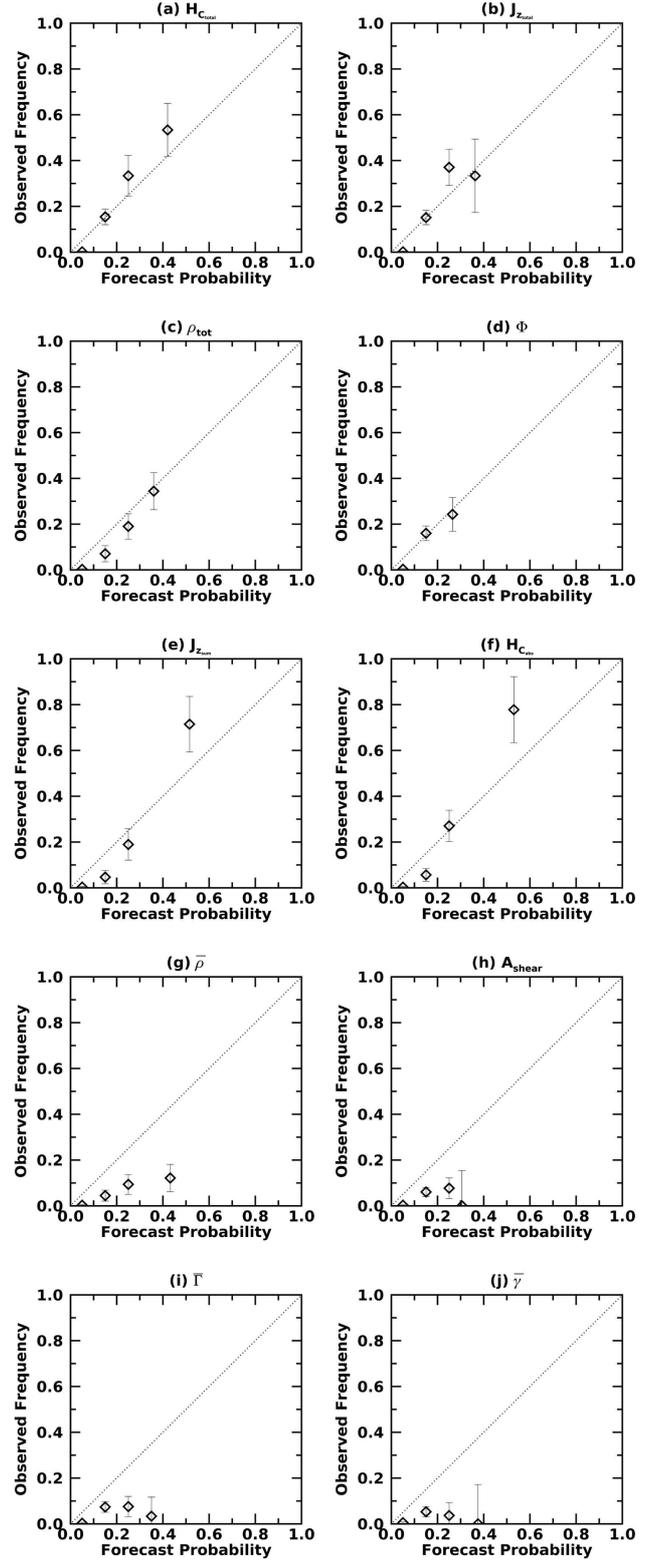}}
\caption{Observed flaring rates versus the predicted rates for major flares in log-log scale. The uncertainties of observed rates for each group are represented by error bars. Perfect reliability occurs when all points lie on the $x = y$ line. Root mean squre error (RMSE) between predicted rates and observed ones are as follows: (a) 0.07, (b) 0.07, (c) 0.06, (d) 0.03, (e) 0.12, (f) 0.13, (g) 0.18, (h) 0.18, (i) 0.19, and (j) 0.29. \label{fig:f2}}
\end{figure}

Figure \ref{fig:f2} show that reliability plots and their RMSEs (i.e., standard deviation of the residuals from the $y=x$ line) between observed rates and predicted ones. Most of the data points for the total unsigned and signed parameters are relatively adjacent to the diagonal line compared to the mean parameters. The total signed parameters can predict a wider range of probabilities than the other types of parameters. The $\Phi$ gives the best reliability (RMSE = 0.03), but the reliability plots and their RMSEs can depend on the selection of probability bins.

\begin{figure*} \centering
\resizebox{\hsize}{!}{\includegraphics{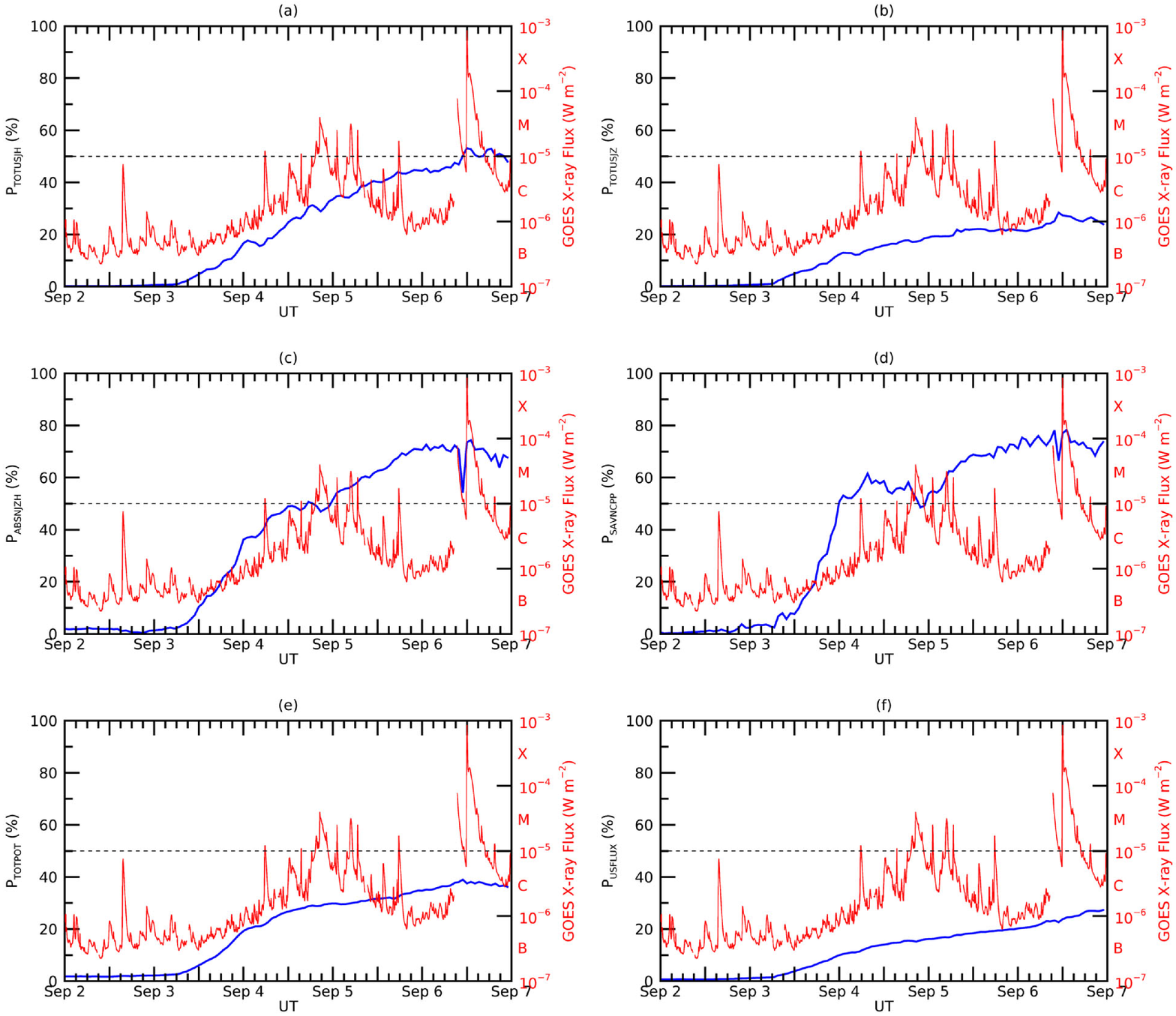}}
\caption{Predicted major flare occurrence probabilities of HARP 7115 (NOAA 12673) using the model of (a) $H_{C\text{,total}}$, (b) $J_{z\text{,total}}$, (c) $H_{{C\text{,abs}}}$, (d) $J_{{z\text{,sum}}}$, (e)  $\rho_{\text{tot}}$ , and (f) $\Phi$ from 2017 September 2 to 6 and \textit{GOES}-15 X-ray flux (5 minute data) in the passbands of 1 - 8 \AA\ (red). These are updated every 1 hour. The dashed line is M-class thresholds.}
\label{fig:f3}
\end{figure*}

As an example, we present the probabilities of HARP 7115 (NOAA 12673) predicted by our models with top six BSS at the cadence of 1 hour from 2017 September 2 to 6 in Figure \ref{fig:f3}. Our forecast models present their own probabilities  of major flare occurrence within a day after the observation time. Our predicted probabilities from six models are compared with GOES-15 X-ray fluxes (5 minute data). During this period, several major flares occurred in HARP 7115 (NOAA 12673). On September 2, probabilities that major flares occur are mostly lower than 20\%. Actually, there is no major flare during that period.
The models of the total signed parameters ($H_{{C\text{,abs}}}$ and $J_{{z\text{,sum}}}$) show a rapid and significant increase of forecast probabilities around September 4 when the first major flare occurred as well as persistence of high probabilities (greater than 50\%) over the next three days. On the other hand, the models of the total unsigned parameters show relatively low probabilities (mostly less than 50\%) with a gradually increasing trend. This suggests that the total signed parameters may help the empirical models in this study to better perform in some cases such as HARP 7115.

Our forecast models are to predict daily flare probability at 00:00 TAI using SDO/HMI observations. However, the models may have an issue on operational forecasting since there is a delay of getting magnetic parameters after the observations. At present, the data that we can get the fastest are available in less than three hours after the observation time. The calculation of the probability from our models needs only a few minutes. Thus, the delay time from observations to forecasts will not exceed 3 hours. For operational forecasting, we can predict daily flaring probability at 00:00 TAI using the observation at 21:00 TAI of the previous day. In order to examine how much this operational model is different with the original model, we investigate the relationship between $J_{z\text{,total}}$ at 21:00 TAI, which has the highest CC between parameter values and flaring rates in Table \ref{tab:tbl2}, and flaring rate for a day (00:00 - 23:59 TAI). In this case, the fitting coefficients, $a$ and $b$, are 1.53 and -21.83, respectively, which corresponds to about 2\% difference with the fitting coefficients of the original model with $J_{z\text{,total}}$ in Table \ref{tab:tbl2}. The predicted probabilities from the operational model are underestimated by 2 - 6.3\% than those from the original model. If we provide the operational forecasting with this model, the forecast has to have this uncertainty together with its original uncertainties.


\subsubsection{Verification measures of Binary}

The probabilistic forecast results have been converted into contingency tables by using probability thresholds \citep{Colak09, Crown12, Bloomfield12}. A contingency table consists of four components: the number of TP (flare event predicted and occurred), FN (no flare event predicted and flare occurred), FP (flare event predicted and did not occur), and TN (no flare event predicted and did not occur) as shown in Table \ref{tab:tbl4}. Verification measures for binary forecasts are obtained by combining these four components of the contingency table. We build contingency tables using probability thresholds and calculate six measures.

\begin{table}[]
\centering
\caption{Contingency table}
\label{tab:tbl4}
\begin{tabular}{ccc}
\hline
\hline
               & \multicolumn{2}{c}{Forecast} \\
Observation & Flare       & No flare       \\ \hline
Yes            & TP          & FN             \\
No             & FP          & TN             \\ \hline
\end{tabular}
\end{table}

In this study, we consider the following verification measures: the proportion correct (PC), the probability of detection (POD), the false alarm ratio (FAR), the Heidke skill score (HSS), the true skill statistic (TSS), and the symmetric extremal dependence index (SEDI). The first five measures have been widely used for the evaluation of flare forecasts, and the last one has been rarely used but is meaningful for validating forecast of rare events such as major flares \citep{Ferro11, Kubo17}. These six performance measures are described in Table \ref{tab:tbl5}.

\begin{table}[]
\caption{Verification Measures}
\label{tab:tbl5}
\centering
\begin{tabular}{ccc}
\hline
\hline
Measure & Equation & Perfect \\ \hline
PC & $\frac{\text{TP} + \text{TN}}{\text{TP} + \text{FN} + \text{FP} + \text{TN}}$ & 1 \\
POD & $\frac{\text{TP}}{\text{TP} + \text{FN}}$ & 1 \\
FAR & $\frac{\text{FP}}{\text{TP} + \text{FP}}$ & 0 \\
TSS & POD - POFD & 1 \\
HSS & $\frac{2[(\text{TP} \times \text{TN}) - (\text{FN} \times \text{FP})]}{(\text{TP} + \text{FN})(\text{FN} + \text{TN}) + (\text{TP} + \text{FP})(\text{FP} + \text{TN})}$ & 1 \\
SEDI & $\frac{(\log (\text{POFD}) - \log (\text{POD}) - \log (1 - \text{POFD}) + \log (1 - \text{POD})}{(\log (\text{POFD}) + \log (\text{POD}) + \log (1 - \text{POFD}) + \log (1 - \text{POD})}$ & 1 \\ \hline
\end{tabular}
\tabnote{A proportion correct (PC) called accuracy is the measure what fraction of the forecast were correct. A quantify what fraction of the flare event observed were correctly forecast is a probability of detection (POD) or hit rate. POD should be used in conjunction with a false alarm ratio (FAR), which is the score what fraction of the predicted flare event actually did not observe. Probability of false detection (POFD) which is the quantity what fraction of the non flare observation were incorrectly forecast to occur. Following two scores are obtained using all components of the contingency table. A true skill statistic (TSS) is a discriminant how well did the forecast discriminate the flare event from the non flare event. Heidke skill score (HSS) represents the accuracy of the forecast relative to that of random chance. Recently, the symmetric extremal dependence index (SEDI) is proposed by \citep{Ferro11}. This skill score is for forecasting rare events.}
\end{table}

We determine the values of six measures and their thresholds to achieve optimum TSS and HSS in contingency tables, which are summarized with their uncertainties in Table \ref{tab:tbl6} and \ref{tab:tbl7}, respectively. In view of the optimized TSS, the total unsigned parameters ($H_{C\text{,total}}$, $\rho_{\text{tot}}$, $J_{z\text{,total}}$, and $\Phi$) achieved higher values of verification measures than the total signed ($H_{{C\text{,abs}}}$ and $J_{{z\text{,sum}}}$) and mean parameters ($\overline{\rho}$, $A_{\text{shear}}$, $\overline{\Gamma}$, and $\overline{\gamma}$). $H_{C\text{,total}}$ is one of the parameters having the highest TSS, which is consistent with the highest \textit{F}-score in the classification between flaring events and non-flaring events \citep{Bobra15}. The rank order of TSS in view of the optimized TSS is also consistent with that of \textit{F}-score from \citet{Bobra15}. When the probability threshold is small, TSS values are very high but FAR values are also high. In view of the optimized HSS, most of total unsigned parameters also have higher value than the signed and mean parameters ($H_{C\text{,total}} > H_{{C\text{,abs}}}$, $J_{z\text{,total}}$ $\approx$ $J_{{z\text{,sum}}}$, and $\rho_{\text{tot}} > \overline{\rho}$). \citet{Bloomfield12} suggested that TSS is optimized when FN/FP is similar to the ratio of flare event number to non-flare event number, and the HSS is optimized when FN/FP $\approx$ 1. The thresholds of the optimum TSS and HSS proposed by \citet{Bloomfield12} are slightly different (from 0 to 0.03) with the thresholds of the optimum TSS and HSS in our models.

\begin{table*}[h]
\caption{Verification measure values and probability thresholds chosen to achieve the optimum true skill statistic (TSS) from ten SHARP magnetic parameters}
\centering
\label{tab:tbl6}
\centering
\resizebox{\textwidth}{!}{
\begin{tabular}{cccccccccccc}
\hline
\hline
Parameter & Threshold & PC & POD &  FAR & TSS & HSS & SEDI \\ \hline
$H_{C\text{,total}}$ & 0.04 & $0.93 \pm 0.002$ & $0.97 \pm 0.003$ & $0.89 \pm 0.004$ & $0.91 \pm 0.004$ & $0.18 \pm 0.006$ & $0.97 \pm 0.0007$ \\
$\rho_{\text{tot}}$ & 0.055 & $0.93 \pm 0.0008$ & $0.97 \pm 0.003$ & $0.9 \pm 0.002$ & $0.91 \pm 0.003$ & $0.18 \pm 0.003$ & $0.97 \pm 0.0004$ \\
$J_{z\text{,total}}$ & 0.045 & $0.93 \pm 0.001$ & $0.97 \pm 0.003$ & $0.9 \pm 0.002$ & $0.91 \pm 0.003$ & $0.18 \pm 0.004$ & $0.96 \pm 0.0005$ \\
$\Phi$ & 0.035 & $0.9 \pm 0.006$ & $0.97 \pm 0.005$ & $0.93 \pm 0.006$ & $0.88 \pm 0.005$ & $0.13 \pm 0.01$ & $0.96 \pm 0.001$ \\
$H_{{C\text{,abs}}}$ & 0.07 & $0.93 \pm 0.001$ & $0.87 \pm 0.007$ & $0.91 \pm 0.003$ & $0.8 \pm 0.007$ & $0.16 \pm 0.004$ & $0.91 \pm 0.004$ \\
$J_{{z\text{,sum}}}$ & 0.03 & $0.84 \pm 0.001$ & $0.95 \pm 0.004$ & $0.95 \pm 0.001$ & $0.79 \pm 0.005$ & $0.07 \pm 0.002$ & $0.91 \pm 0.003$ \\
$\overline{\rho}$ & 0.03 & $0.8 \pm 0.0007$ & $0.97 \pm 0.003$ & $0.96 \pm 0.0008$ & $0.78 \pm 0.003$ & $0.06 \pm 0.001$ & $0.91 \pm 0.0007$ \\
$A_{\text{shear}}$ & 0.015 & $0.74 \pm 0.006$ & $0.95 \pm 0.006$ & $0.97 \pm 0.0006$ & $0.68 \pm 0.003$ & $0.04 \pm 0.001$ & $0.84 \pm 0.003$ \\
$\overline{\Gamma}$ & 0.015 & $0.7 \pm 0.037$ & $0.95 \pm 0.035$ & $0.98 \pm 0.004$ & $0.65 \pm 0.007$ & $0.03 \pm 0.008$ & $0.82 \pm 0.007$ \\
$\overline{\gamma}$ & 0.025 & $0.77 \pm 0.003$ & $0.84 \pm 0.008$ & $0.97 \pm 0.0009$ & $0.61 \pm 0.008$ & $0.04 \pm 0.001$ & $0.77 \pm 0.007$ \\ \hline
\end{tabular}
}
\end{table*}

\begin{table*}[h]
\caption{Verification measure values and probability thresholds chosen to achieve the optimum Heidke skill score (HSS) from ten SHARP magnetic parameters}
\centering
\label{tab:tbl7}
\centering
\resizebox{\textwidth}{!}{
\begin{tabular}{cccccccccccc}
\hline
\hline
Parameter & Threshold & PC & POD &  FAR & TSS & HSS & SEDI \\ \hline
$H_{C\text{,total}}$ & 0.195 & $0.99 \pm 0.0001$ & $0.45 \pm 0.01$ & $0.55 \pm 0.01$ & $0.44 \pm 0.01$ & $0.44 \pm 0.009$ & $0.76 \pm 0.007$ \\
$H_{{C\text{,abs}}}$ & 0.21 & $0.99 \pm 0.0008$ & $0.45 \pm 0.03$ & $0.59 \pm 0.03$ & $0.44 \pm 0.03$ & $0.43 \pm 0.009$ & $0.76 \pm 0.02$ \\
$J_{z\text{,total}}$ & 0.15 & $0.99 \pm 0.0004$ & $0.61 \pm 0.02$ & $0.68 \pm 0.007$ & $0.59 \pm 0.02$ & $0.41 \pm 0.008$ & $0.83 \pm 0.009$ \\
$J_{{z\text{,sum}}}$ & 0.22 & $0.99 \pm 0.0003$ & $0.45 \pm 0.02$ & $0.61 \pm 0.03$ & $0.44 \pm 0.02$ & $0.41 \pm 0.009$ & $0.75 \pm 0.008$ \\
$\rho_{\text{tot}}$ & 0.205 & $0.98 \pm 0.001$ & $0.58 \pm 0.05$ & $0.75 \pm 0.03$ & $0.57 \pm 0.05$ & $0.34 \pm 0.007$ & $0.8 \pm 0.02$ \\
$\Phi$ & 0.135 & $0.98 \pm 0.001$ & $0.55 \pm 0.04$ & $0.77 \pm 0.009$ & $0.54 \pm 0.04$ & $0.32 \pm 0.006$ & $0.78 \pm 0.02$ \\
$\overline{\rho}$ & 0.185 & $0.96 \pm 0.007$ & $0.53 \pm 0.09$ & $0.89 \pm 0.009$ & $0.49 \pm 0.08$ & $0.17 \pm 0.004$ & $0.71 \pm 0.06$ \\
$A_{\text{shear}}$ & 0.12 & $0.94 \pm 0.003$ & $0.58 \pm 0.03$ & $0.93 \pm 0.002$ & $0.52 \pm 0.03$ & $0.11 \pm 0.003$ & $0.71 \pm 0.02$ \\
$\overline{\Gamma}$ & 0.1 & $0.93 \pm 0.02$ & $0.68 \pm 0.17$ & $0.93 \pm 0.005$ & $0.61 \pm 0.15$ & $0.12 \pm 0.003$ & $0.78 \pm 0.12$ \\
$\overline{\gamma}$ & 0.12 & $0.95 \pm 0.003$ & $0.37 \pm 0.02$ & $0.94 \pm 0.002$ & $0.32 \pm 0.02$ & $0.09 \pm 0.003$ & $0.54 \pm 0.02$ \\ \hline
\end{tabular}
}
\end{table*}

\begin{figure*}\centering
\includegraphics[width=10cm]{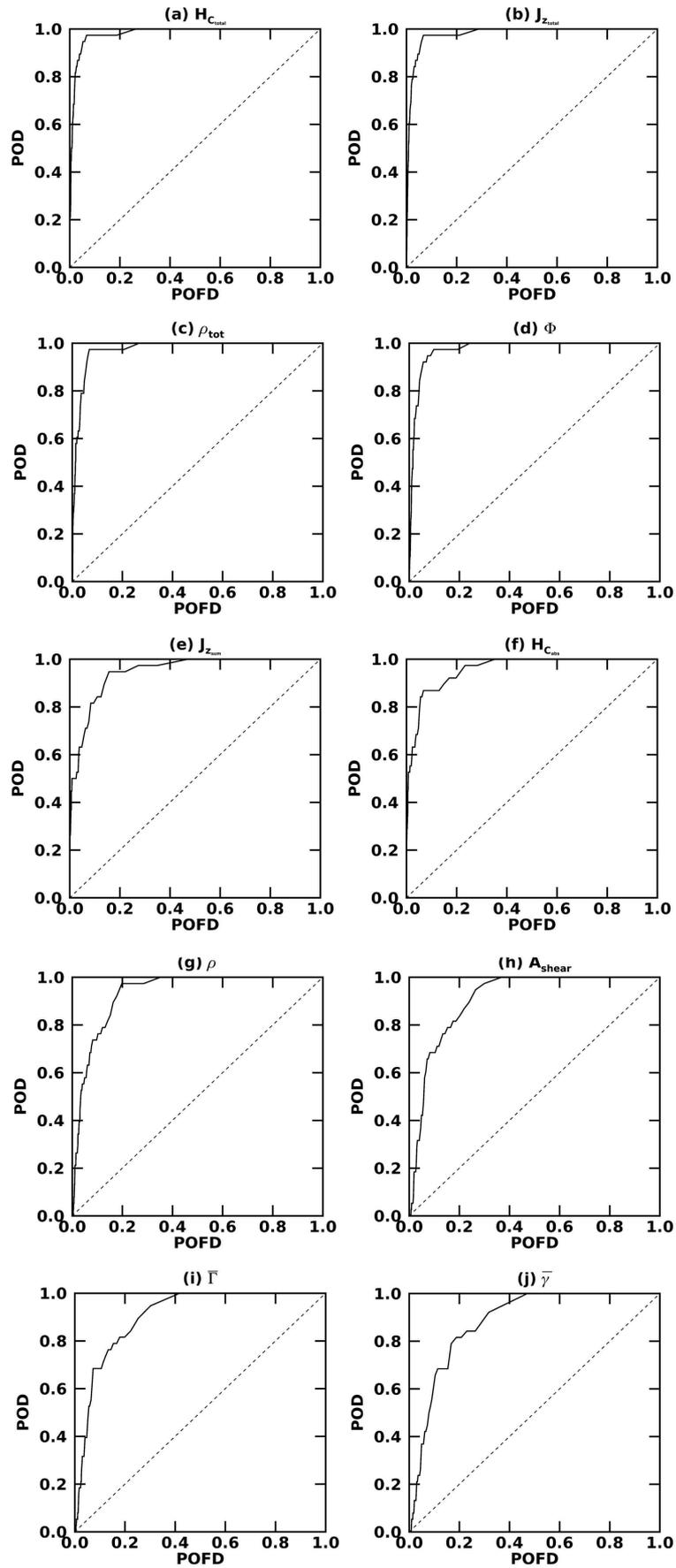}
\caption{Receiver operating characteristic (ROC) curve in which probability of detection (POD) is plotted as a function of probability of false detection (POFD). When the perfect forecast occurs, the curve travels from bottom left to top left of diagram, then across to top right of diagram.}
\label{fig:f4}
\end{figure*}

Another interesting measure is a relative operating characteristic (ROC) curve in which POD is plotted as a function of POFD and a ROC area which is a area under the ROC curve. When the perfect forecast occurs, the curve travels from bottom left to top left of diagram, then across to top right of diagram. In this case, the ROC area is 1. ROC areas under the curve of ten parameters with their uncertainties are as follows: $0.98 \pm 0.001$ from $H_{C\text{,total}}$, $0.98 \pm 0.001$ from $J_{z\text{,total}}$, $0.97 \pm 0.001$ from $\rho_{\text{tot}}$, $0.97 \pm 0.001$ from $\Phi$, $0.95 \pm 0.002$ from $J_{{z\text{,sum}}}$, $0.96 \pm 0.001$ from $H_{{C\text{,abs}}}$, $0.93 \pm 0.001$ from $\overline{\rho}$, $0.91 \pm 0.002$ from $A_{\text{shear}}$, $0.9 \pm 0.002$ from $\overline{\Gamma}$, and $0.88 \pm 0.002$ from $\overline{\gamma}$. As shown in Figure \ref{fig:f4}, the total unsigned parameters have higher ROC areas than the total signed parameters and the mean parameters, which gives the better performance in terms of the ROC curve measure. The difference in ROC areas is very small between all of the total unsigned parameters.

\subsubsection{Uncertainties of verification measures}

Several forecast studies have mentioned the necessity of uncertainties for forecast verification measures \citep{Barnes16, Kubo17, Leka18} because the verification measures are calculated from finite number of samples. To estimate uncertainties of verification measures, we use a bootstrap method, which accounts for random errors using resampling. We consider the size of resampled data which is the difference between the number of non-events and events. We make a resampled data with size $=4648$ by picking randomly. Then, we calculate the verification measures from the resampled data. This process iterates 1000 times and the uncertainties are estimated by the standard deviation of the resampled values of verification measures in Table \ref{tab:tbl6} and \ref{tab:tbl7}.


\section{Summary and Discussion} \label{sec:summary}

We have presented the forecasting models of major flare probability based on the power law relationships between ten magnetic parameters and major flare occurrence rates. The magnetic parameters calculated from \textit{SDO}/HMI vector magnetic fields are used. The data are taken from May 2010 to April 2018 and divided into two sets (training of 11,040 HARPs and test of 4,724 HARPs) in chronological order, which is proper for forecast purpose. All values of the magnetic parameters are divided into 50 subgroups to estimate corresponding flare occurrence rates. From this, we considered the power law relationships between magnetic parameters and flaring rates.

The major results of this study are summarized as follows. First, major flare occurrence rates are well correlated with ten magnetic parameters (CC $\geq$ 0.86). Second, the logarithmic values of flaring rates are well approximated by the linear equations.
Third, the total unsigned parameters achieved relatively higher values of the optimized TSS and HSS than the total signed and mean parameters. Among the total unsigned parameters, $H_{C\text{,total}}$, $\rho_{\text{tot}}$, and $J_{z\text{,total}}$ are also the highest ranked of univariate \textit{F}-scores in \citet{Bobra15}. Our results are well consistent with \citet{Toriumi17} who found that $H_{C\text{,total}}$, $\rho_{\text{tot}}$, and $J_{z\text{,total}}$ are stronger proportional to magnetic free energy, which is calculated from 3D reconstructed magnetic fields, than the other parameters.

Although we calculate various verification measures of ten magnetic parameters, it is difficult to conclude that what parameter is outperformed than the others. Most of parameters have high values of measures with only small differences. The result also depends on verification, where the total unsigned parameters have slightly better performance than the total signed parameters in terms of optimized TSS and HSS, but the total signed parameters can predict higher probabilities than the total unsigned parameters in practical forecasts for given AR. The performances of converted binary forecasts depend on the probability thresholds. When the probability threshold is small, TSS are very high but FAR are also high. On the other hand, all verification measure values are moderate in case of optimized HSS. Thus, the decision-maker or user may select a proper threshold or model for their own purpose.

Our parameters that give a relatively good performance are conventionally considered to non-potential parameters. This result supports the importance of non-potential magnetic fields in ARs (e.g., \citealt{Low94, Canfield99, Schrijver09, Jing12, Zhang16}).
\citet{Bao99} have found that the time variations of current helicity in the highly flaring active regions are more significant than those of the poorly flaring active regions. The magnetic free energy plays an important role in producing major flares, which has been already well known \citep{Canfield99, Moore01}. \citet{Ji03} revealed a quite high correlation between vertical current and flares.
\citet{Liu16} suggested that the total unsigned vertical current and the photospheric magnetic free energy should be responsible for flare productivity.
Our results together with the previous results demonstrate an importance of total quantity of non-potential magnetic properties for flare forecasting \citep{Welsch09, Bobra15, Liu17, Toriumi17}.


\acknowledgments

We acknowledge the referees for giving us useful comments. This work was supported by the BK21 plus program through the National Research Foundation (NRF) funded by the Ministry of Education of Korea, the Basic Science Research Program through the NRF funded by the Ministry of Education (NRF-2016R1A2B4013131, NRF-2016R1A6A3A11932534, NRF-2019R1A2C1002634), NRF of Korea Grant funded by the Korean Government (NRF-2013M1A3A3A02042232), the Korea Astronomy and Space Science Institute under the R\&D program supervised by the Ministry of Science, ICT and Future Planning, the Korea Astronomy and Space Science Institute under the R\&D program ‘Development of a Solar Coronagraph on International Space Station (Project No. 2019-1-850-02)’ supervised by the Ministry of Science, ICT and Future Planning, and Institute for Information \& communications Technology Promotion (IITP) grant funded by the Korea government (MSIP) (2018-0-01422, Study on analysis and prediction technique of solar flares). The data used here are courtesy of NASA/\textit{SDO} and the HMI science team, as well as the \textit{GOES} team. 


\end{document}